\documentclass[prx,twocolumn,showpacs,preprintnumbers,amsmath,amssymb,citeautoscript,superscriptaddress]{revtex4-2}

\usepackage{graphicx}
\usepackage{amssymb, amsfonts, amsmath}
\usepackage{here}
\usepackage{dcolumn}
\usepackage{bm}
\usepackage{color}
\usepackage{here}
\usepackage{xr}
\usepackage{textcomp}

\makeatletter

\newlength{\figwidth}
\newcommand{\FigCap}[1]{(#1)}	

\newcommand{\ts}{$T_{{\rm s}}$}
\newcommand{\tc}{$T_{{\rm c}}$}

\newcommand{\tw}{$T_{0}$}

\newcommand{\FeSeTe}{FeSe$_{1-x}$Te$_{x}$}
\newcommand{\FeSeS}{FeSe$_{1-x}$S$_{x}$}

\makeatother

\begin{document}

\title{Pure nematic quantum critical point accompanied by a superconducting dome}

\author{K. Ishida}\email{kousuke.ishida@cpfs.mpg.de}
\altaffiliation[Present address: ]
{Max Planck Institute for Chemical Physics of Solids, N\"{o}thnitzer Stra{\ss}e 40, 01187 Dresden, Germany.}
\author{Y. Onishi}
\altaffiliation[Present address: ]
{Department of Applied Physics, The University of Tokyo, Hongo, Tokyo 113-8656, Japan.}
\author{M. Tsujii}
\author{K. Mukasa} 
\author{M. Qiu}
\author{M. Saito}
\author{\\Y. Sugimura}
\author{K. Matsuura}
\altaffiliation[Present address: ]
{Research Center for Advanced Science and Technology (RCAST), University of Tokyo, Meguro-ku, Tokyo 153-8904, Japan.}
\author{Y. Mizukami}
\author{K. Hashimoto}
\author{T. Shibauchi}\email{shibauchi@k.u-tokyo.ac.jp}
\affiliation{%
 Department of Advanced Materials Science, University of Tokyo, Kashiwa, Chiba 277-8561, Japan.
}%

\date{\today}

\begin{abstract}
	When a symmetry-breaking phase of matter is suppressed to a quantum critical point (QCP) at absolute zero, quantum-mechanical fluctuations proliferate. 
	Such fluctuations can lead to unconventional superconductivity, as evidenced by the superconducting domes often found near magnetic QCPs in correlated materials. 
	However, it remains unclear whether this superconductivity mechanism holds for QCPs of the electronic nematic phase, characterized by rotational symmetry breaking. 
	Here, we demonstrate from systematic elastoresistivity measurements that nonmagnetic \FeSeTe \ exhibits an electronic nematic QCP showing diverging nematic susceptibility. 
	This finding establishes two nematic QCPs in FeSe-based superconductors with contrasting accompanying phase diagrams. 
	In \FeSeTe, a superconducting dome is centered at the QCP, whereas \FeSeS \ shows no QCP-associated enhancement of superconductivity. 
	We find that this difference is related to the relative strength of nematic and spin fluctuations. 
	Our results in \FeSeTe \ present the first case in support of the superconducting dome being associated with the pure nematic QCP.

\end{abstract}

\maketitle

\section{INTRODUCTION}
In unconventional superconductors, the interplay between superconductivity and a quantum critical point (QCP), defined as the point of continuous phase transition at absolute zero temperature, has been one of the central topics for decades.
At the QCP, the ground state becomes a quantum superposition of the ordered and disordered states, giving rise to enhanced quantum mechanical fluctuations 
\cite{sachdev2011quantum}.
These enhanced quantum fluctuations couple to the low-energy quasiparticle excitations near the Fermi energy, which causes the non-Fermi liquid power-law behavior of the physical quantities and sometimes leads to the formation of Cooper pairs.
In particular, the focus of interest has been on the antiferromagnetic QCPs found in many classes of unconventional superconductors such as copper oxides, iron pnictides, and heavy-fermion materials \cite{keimer2015quantum,shibauchi2014quantum,mathur1998magnetically}. 
The phase diagrams of these materials show a dome-shaped superconducting phase near the vanishing point of the antiferromagnetic phase, which implies that strong spin fluctuations near the antiferromagnetic QCP can mediate the superconductivity with high critical temperature $T_{\rm c}$ \cite{moriya2003antiferromagnetic}. 

In recent years, however, several kinds of unconventional superconductors have been also found to exhibit electronic nematic orders, which break rotational symmetry of the underlying lattice, close to the superconducting dome \cite{fradkin2010nematic,ishida2020divergent,ronning2017electronic}.
Most electronic nematic orders are ferroic and do not break additional translational symmetry, and thus nematic fluctuations have a peak at the zero wavevector $\bm{q} \sim \bm{0}$, in contrast to the antiferromagnetic fluctuations which are strong at finite wavevectors such as $\bm{q} \sim (\pm\pi,\pm\pi)$ in cuprates.
It is then important to investigate whether such nematic fluctuations can promote unconventional superconductivity or not. Recent theoretical studies have pointed out that they can enhance the critical temperature \tc \  \cite{lederer2015enhancement,maier2014pairing}.
In real materials, however, the nematic order often coexists with other competing spin or charge orders, which makes it challenging to experimentally establish that nematic fluctuations themselves can strengthen the superconducting pairing.

Iron-chalcogenide superconductor FeSe is an ideal system to address this issue.
This compound exhibits an electronic nematic phase below the tetragonal-to-orthorhombic structural transition temperature \ts \ $\sim 90$\,K, but unlike most iron-based superconductors, it does not show any long-range magnetic order down to zero temperature \cite{shibauchi2020exotic}.
By applying hydrostatic pressure, \ts \ of FeSe is rapidly suppressed, but before the vanishing of nematic order, a pressure-induced antiferromagnetic order sets in. This magnetic phase shows a dome shape in the pressure phase diagram, whereas \tc \ exhibits a four-fold increase from 9\,K at ambient pressure up to $\sim 37$\,K when the antiferromagnetism is suppressed at high pressure \cite{sun2016dome}.
In contrast, the isovalent S substitution for Se can suppress \ts \ to zero temperature without stabilizing the magnetic order, but \tc \ is found to show an abrupt decrease across the endpoint of the nematic phase \cite{reiss2017suppression}. 
Thus, experimentally, there is no evidence in S-substituted FeSe that nematic fluctuations enhanced at the endpoint of \ts \ promote superconductivity.

\begin{figure}[t]
	\centering
	\includegraphics[width=1\linewidth]{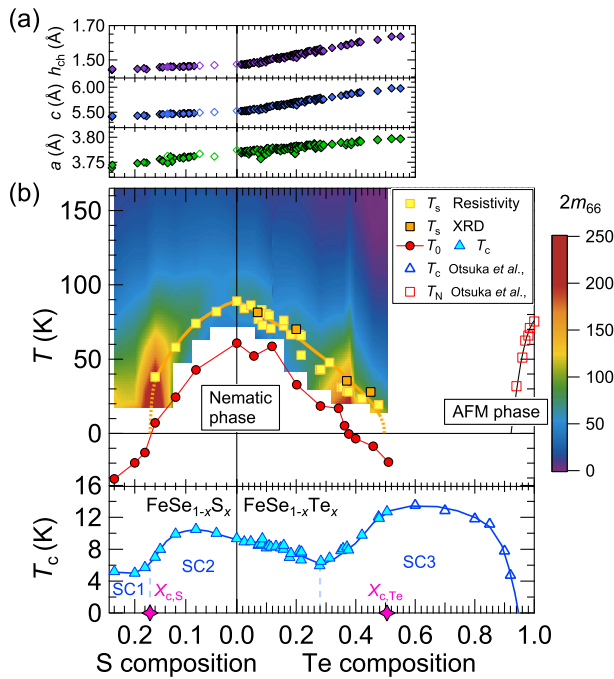}
	\caption{Nonmagnetic nematic quantum critical points in FeSe-based superconductors.
	\FigCap{a}, S and Te composition dependence of the $a$-axis length (bottom), $c$-axis length (middle), and the height of the chalcogen from Fe-plane (top). Open diamonds and closed pentagons are from refs.\,\cite{matsuura2017maximizing} and \cite{Mukasa2020high}, respectively. 
	\FigCap{b}, Combined phase diagram of \FeSeS \ and \FeSeTe, which includes \ts \ determined by the resistivity (yellow squares) and X-ray diffraction experiments (orange squares) \cite{Mukasa2020high}, \tc \ (blue triangles), antiferromagnetic transition temperature $T_{\rm N}$ (red squares) \cite{otsuka2019incoherent}, and \tw \ obtained from the analysis of $2m_{66}$ data. The data of closed symbols are from the single crystals grown by the chemical vapor transport technique \cite{Mukasa2020high}, and the data of open symbols in highly Te substituted region are taken from the study of single crystals synthesized by the flux method \cite{otsuka2019incoherent}. The magnitude of $2m_{66}$ is shown as a color plot. \tc \ has a two-dome structure, which contains two nematic quantum critical points ($x_{\rm c,S}$ and $x_{\rm c,Te}$) and the superconducting state can be separated to three regions (SC1, SC2, and SC3), whose properties are considered to be different.          
	}
	\label{PD}
\end{figure}

The Se site of FeSe can also be substituted by isovalent Te. Previously, the single-crystal studies of \FeSeTe \ are almost limited to the high Te-composition side ($x\gtrsim0.5$), which have shown that FeSe$_{0.5}$Te$_{0.5}$ does not exhibit nematic order with \tc \ as high as 14\,K \cite{Li2009,Sales2009}.
Recent advances on the single-crystal growth of \FeSeTe \ by the flux method under the temperature gradient conditions \cite{terao2019superconducting} and the chemical vapor transport (CVT) technique \cite{Mukasa2020high} have enabled to trace systematically the nematic and superconducting transition temperatures with Te substitution, by overcoming the phase separation issue previously reported for $0.10 \lesssim x \lesssim 0.30$ \cite{fang2008superconductivity}.
It has been found that \ts \ of the CVT-grown single crystals is monotonically suppressed with increasing Te concentrations and disappears around $x = 0.50$, whereas \tc \ first decreases and reaches its minimum at $x \sim 0.30$, and then turns to increase \cite{Mukasa2020high}, as reproduced in the right panel of Fig.\,\ref{PD}\FigCap{b}.

The increase in \tc \ toward the endpoint of \ts \ points to a potential link between suppressed nematicity and enhanced superconductivity in \FeSeTe.
However, this increasing trend of \tc \ with vanishing \ts \ stands in marked contrast to the phase diagram of \FeSeS, which raises the fundamental question on the origin of this difference.
To discuss the above issues, it is essential to clarify how nematic fluctuations evolve with Te substitutions compared with the S-substitution case and whether we have a nematic QCP in the phase diagram of \FeSeTe.
Here, by performing systematic elastoresistivity measurements on \FeSeTe \ single crystals to quantify the nematic susceptibility, we demonstrate that \FeSeTe \ is an unprecedented system whose nematic QCP lies near the center of the superconducting dome in isolation to any other long-range orders.
Comparisons between the results in \FeSeTe\ and  \FeSeS \ imply that the dominance of nematic fluctuations over antiferromagnetic fluctuations is the key to enhancing \tc \ around the nonmagnetic nematic QCP. 

\section{Results}


\begin{figure}[t]
	\centering
	\includegraphics[width=1\linewidth]{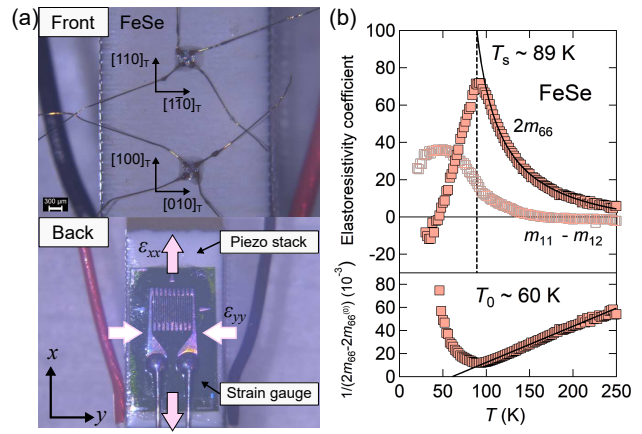}
	\caption{$B_{1g}$ and $B_{2g}$ nematic susceptibilities of FeSe measured by elastoresistivity technique.
	\FigCap{a}, Photographs of the elastoresistance measurement setup. 
	The square-shaped samples with electric contacts on four corners were directly glued on the piezo stack. 
	On the back side, the strain gauge was attached to measure the amount of the applied strain. 
	For the $B_{2g}$ and $B_{1g}$ nematic susceptibility measurements, the samples are aligned along the $[110]_{\rm T}$ and $[100]_{\rm T}$ directions, respectively.
	\FigCap{b}, Top panel shows the temperature dependence of the two elastoresistivity coefficients of FeSe. The black line represents the Curie-Weiss fit for $2m_{66}(T)$. Bottom panel displays the inverse of $2m_{66}-2m_{66}^{(0)}$, where $2m_{66}^{(0)}$ is determined by the Curie-Weiss fitting.}
	\label{ER}
\end{figure}

Nematic order is characterized by rotational symmetry breaking, and thus its order parameter can be expressed by the anisotropy of physical quantities such as electrical resistivity \cite{fradkin2010nematic}.
Since the uniaxial strain works as a conjugate field to the nematic order parameter, the nematic susceptibility above the transition temperature \ts\ can be obtained from the electronic anisotropy induced by the strain applied to the system as a perturbation.  
In our elastoresistivity measurements, we assume the in-plane resistivity anisotropy as an order parameter of the nematic phase, and the anisotropic biaxial strain is applied using the piezoelectric device (see Methods) \cite{chu2012divergent}.
As shown in Fig.\,\ref{ER}\FigCap{a}, for the resistivity measurements along two directions on a single sample, we apply the Montgomery method to the square-shaped crystals.
The samples are directly glued on the surface of the piezo stacks, and the strain is controlled by applying the voltage to the device and monitored by the strain gauge attached on the other side.

In the tetragonal FeSe-based materials with $D_{4h}$ point group, there are two candidates for the in-plane nematic order.
One is along the adjacent Fe-chalcogen direction with $B_{1g}$ irreducible representation, and the other is along the Fe-Fe direction with $B_{2g}$ symmetry (here we use the experimental 2-Fe unit cell notation).
Nematic susceptibility for each symmetry channel can be measured by applying the strain along its corresponding direction, and by using the elastoresistivity tensor defined as $m_{ij}=(\Delta \rho/\rho)_i/\varepsilon_j$, where $(\Delta \rho/\rho)_i$ is the relative change of resistivity against the strain $\varepsilon_j$ with subscript $i$ and $j$ represented by the Voigt notation ($1 = xx, 2 = yy, 3 = zz, 4 = yz, 5 = zx, 6 = xy$), the $B_{1g}$ and $B_{2g}$ components can be expressed as $m_{11}-m_{12}$ and 2$m_{66}$, respectively \cite{kuo2016ubiquitous}.


Figure\,\ref{ER}\FigCap{b} shows the temperature dependence of the two nematic susceptibilities for FeSe single crystals.
Above \ts, 2$m_{66}$ displays a strong temperature evolution with much larger magnitude compared to $m_{11}-m_{12}$, confirming the $B_{2g}$ Ising nematic order of FeSe.     
Furthermore, in the disordered state above \ts, 2$m_{66}$ obeys the Curie-Weiss law
\begin{equation}
2m_{66}(T)=\frac{a}{T-T_0}+2m_{66}^{(0)},	
\end{equation}
where $a$ and $2m_{66}^{(0)}$ are temperature-independent constants.
The Curie-Weiss temperature $T_0$ gives the bare nematic transition temperature in the absence of nemato-elastic coupling in the system.
However, the presence of finite coupling shifts the thermodynamic nematic transition from \tw \ to \ts ($>T_0$). 
Inside the ordered phase below \ts, $2m_{66}(T)$ no longer follows the Curie-Weiss law, as shown in Fig.\,\ref{ER}\FigCap{b}.
The Curie-Weiss temperature dependence of the nematic susceptibility above \ts\ in FeSe has also been reported in the previous measurements of the elastoresistivity, Young modulus, and Raman scattering \cite{hosoi2016nematic,tanatar2016origin,bohmer2015origin, massat2016charge}.
We note that this mean-field type divergent behavior in FeSe without magnetism can be explained by the theory based on the orbital-driven nematicity \cite{yamakawa2016nematicity}. 

\begin{figure*}[htbp]
	\centering
	\includegraphics[width=1\linewidth]{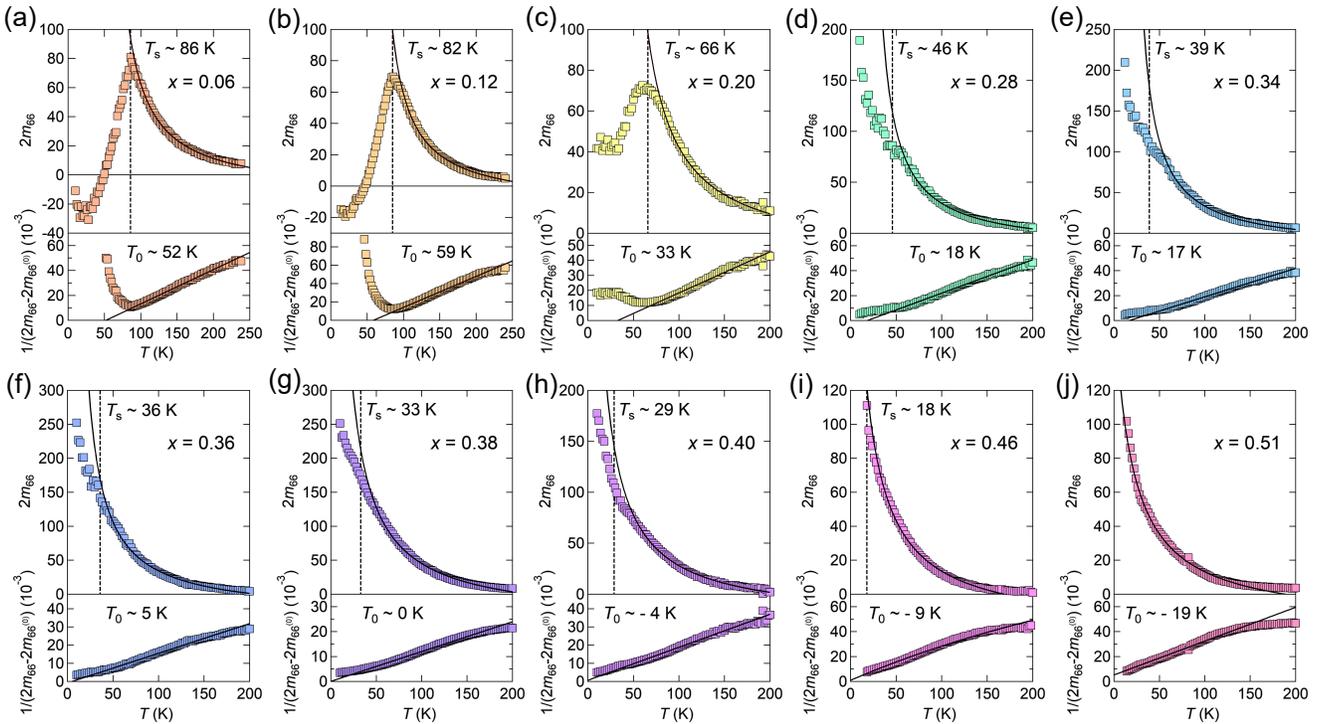}
	\caption{Evolution of elastoresistivity coefficients $2m_{66}$ in \FeSeTe.
	\FigCap{a-j}, Temperature dependence of $B_{2g}$ nematic susceptibilities (top panels) and their Curie-Weiss analyses (bottom panels) for $x =0.06$ \FigCap{a}, $0.12$ \FigCap{b}, $0.20$ \FigCap{c}, $0.28$ \FigCap{d}, $0.34$ \FigCap{e}, $0.36$ \FigCap{f}, $0.38$ \FigCap{g}, $0.40$ \FigCap{h}, $0.46$ \FigCap{i}, and $0.51$ \FigCap{j}.
	In each figure, their structural transition temperatures \ts \ are shown by the vertical dashed line except for $x =0.51$ with no structural transition.
	Black lines represent the Curie-Weiss fitting, and the obtained Curie-Weiss temperatures \tw \ are indicated in the bottom panels.}
	\label{NS}
\end{figure*}

Next, we discuss the evolution of nematic susceptibility with Te substitution.
Single crystals of \FeSeTe \ used in this study were grown by the CVT technique, which can tune the lattice parameters continuously (Fig.\,\ref{PD}\FigCap{a}).
The CVT-grown crystals show homogeneous distributions of Te ions (Fig.\,\ref{figS1}) with no resistivity upturn \cite{Mukasa2020high}, which is caused by the localization effects due to excess Fe as reported in crystals synthesized by the Bridgman method for $x\gtrsim 0.5$ \cite{Sun2019,jiang2020nematic}. 
As depicted in Fig.\,\ref{NS}\FigCap{a-j}, our systematic measurements in a wide range of Te composition $0\le x\le 0.51$ reveal that with increasing $x$, the $B_{2g}$ nematic susceptibility 2$m_{66}$ exhibits a continuous evolution with a gradual decrease in \ts.  
For all Te compositions with finite \ts, the temperature dependence of 2$m_{66}$ above \ts \ can be reasonably described by the Curie-Weiss function, evidencing for their continuous nematic transitions. 
The Curie-Weiss temperature dependence can be also seen at low temperatures in the tetragonal $x = 0.51$, but here we find a clear deviation from the Curie-Weiss behavior at high temperatures $T\gtrsim160$\,K, which may be related to the loss of $d_{xy}$ orbital spectral weight \cite{yi2015observation,jiang2020nematic}.
The magnitude of 2$m_{66}$ becomes largest at $x = 0.38$, in which the Curie-Weiss temperature becomes \tw \ $\sim 0$\,K.
For comparison, we have also measured the $B_{1g}$ nematic susceptibility $m_{11}-m_{12}$, which is found to be much less significant than 2$m_{66}$ even in the tetragonal $x = 0.51$ (Fig.\,\ref{figS2}), demonstrating that the large signal in 2$m_{66}$ solely comes from the $B_{2g}$ nematic response covering the entire Te composition range of the present study.
The observed much weaker temperature dependence of $m_{11}-m_{12}$ also highlights that the magnetic interaction connected to the double-stripe magnetism in FeTe, which is parallel to the Fe-Te direction with $B_{1g}$ symmetry, is negligible in this $x$ region \cite{bao2009tunable}.

From the elastoresistivity measurements, we map out the the magnitude of 2$m_{66}$ in the phase diagram of \FeSeTe \ (Fig.\,\ref{PD}\FigCap{b}).  
The Curie-Weiss temperature \tw \ deceases almost monotonically with increasing Te composition, and crosses the zero temperature line around $x = 0.38$, where the magnitude of 2$m_{66}$ is strongly enhanced.
Since there have been no reports for long-range magnetic order up to $x = 0.90$ \cite{otsuka2019incoherent}, the observed diverging $B_{2g}$ nematic susceptibility toward 0\,K evidences that if the electron subsystem were not under the lattice environment, we would have the nonmagnetic nematic QCP around $x = 0.38$.
Note that the intensity of 2$m_{66}$, which measures the dynamic nematic susceptibility, should become strongest at \tw \ $\sim 0$\,K because it sees the scale of bare nematic transition temperature. 
The thermodynamic QCP, in which the continuous electronic nematic transition takes place at zero temperature, is shifted to the end point of \ts \ due to the inevitable finite nemato-elastic coupling.

This can be compared with the $x$ dependence of \tc, combined with the previous reports for $x\gtrsim0.6$ (Fig.\,\ref{PD}\FigCap{b}), which clearly indicates that the nematic QCP in this system locates near the center of the superconducting dome. 
This implies a close correlation between the nematic quantum phase transition and enhanced superconductivity. 
Our results are consistent with the recent study for the Te-rich side using Bridgman crystals, which shows the smooth suppression of 2$m_{66}$ for $x\gtrsim0.5$ \cite{jiang2020nematic}. 

\begin{figure}[t]
	\centering
	\includegraphics[width=0.7\linewidth]{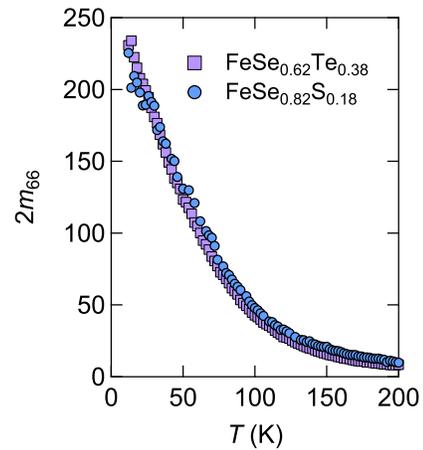}
	\caption{Divergent $B_{2g}$ nematic susceptibilities near the nonmagnetic nematic QCPs of \FeSeTe \ and \FeSeS.
	Temperature dependence of $2m_{66}$ for \FeSeTe \ with $x=0.38$ (purple square) from Fig.\,\ref{NS}\FigCap{g} is compared with that for \FeSeS \ with $x=0.18$ (blue circle) from Fig.\,\ref{figS3}\FigCap{d}.}
	\label{NS_QCP}
\end{figure}

Our results on \FeSeTe \ indicate that the isovalent Te substitution for Se, which may be considered as a negative chemical pressure effect (see Fig.\,\ref{PD}\FigCap{a}), affects the superconductivity in a completely different way from the S substitution corresponding to positive chemical pressure.
To gain more insights into the difference between the Te and S substitution effects, we also performed the elastoresistivity measurements for \FeSeS \ in the same experimental setup (Fig.\,\ref{figS3}), and plotted the intensities of $2m_{66}$ on the same scale with that of \FeSeTe \ in the combined phase diagram shown in Fig.\,\ref{PD}\FigCap{b}.
As previously reported \cite{hosoi2016nematic}, 2$m_{66}$ of \FeSeS \ also follows the Curie-Weiss temperature dependence, and its \tw \ changes sign around $x=0.17$ with a strong enhancement of the magnitude of nematic susceptibility, demonstrating the $B_{2g}$ nematic QCP (see Fig.\,\ref{figS3}).
Furthermore, the singular behavior in 2$m_{66}$ at the QCP of \FeSeS \ is found to be quite similar to that of \FeSeTe\ (Fig.\,\ref{NS_QCP}), suggesting that the underlying nematic quantum critical behavior is essentially the same between the two systems.

\section{Discussion}
Although diverging behavior of 2$m_{66}$ around the nematic QCP of \FeSeS \ is almost identical to that of \FeSeTe, \tc \ of \FeSeS \ exhibits a sudden decrease across the quantum phase transition, which is in sharp contrast to the superconducting dome in \FeSeTe.
The possible origin of this suppression in the tetragonal \FeSeS \ is that the superconducting state (SC1) in the tetragonal phase of \FeSeS \ is quite different from that in the nematic phase (SC2). Indeed, a recent theoretical study has proposed a possible topological transition into the so-called ultranodal pair state with Bogoliubov Fermi surface in the SC1 region \cite{setty2020topological}.
Although this exotic superconducting state remains still elusive, it can naturally account for the large residual density of states inferred from the specific heat and scanning tunneling spectroscopy, which is suddenly appeared in the tetragonal SC1 phase of \FeSeS \  \cite{sato2018abrupt,hanaguri2018two}.

As illustrated in the bottom part of Fig.\,\ref{PD}\FigCap{b}, the dependence of \tc \ on S concentration inside the nematic phase of \FeSeS \ shows a broad peak structure, which is connected continuously to that in \FeSeTe \ across $x=0$ (FeSe), forming a superconducting dome (SC2). In this SC2 region of \FeSeS, several bulk probes and surface sensitive techniques have provided evidence for the anisotropic superconducting gap \cite{shibauchi2020exotic,sato2018abrupt,hanaguri2018two}.
Recent nuclear magnetic resonance (NMR) experiments revealed that antiferromagnetic fluctuations with $(\pi,\pi)$ wavevector are enhanced inside the nematic phase of \FeSeS, which appears to be in correspondence with the \tc \ dome \cite{wiecki2018persistent}. 
Moreover, high-pressure studies in \FeSeS \ demonstrated that \tc \ is enhanced around the endpoints of pressure-induced antiferromagnetic phase \cite{matsuura2017maximizing}.
These results imply the close relationship between antiferromagnetic fluctuations and enhanced superconductivity in the SC2 region.

In \FeSeTe, the $x$ dependence of \tc \ shows a minimum at $x\sim0.30$, above which another superconducting dome emerges around the nematic QCP found in this study. 
This nonmonotonic \tc$(x)$ in \FeSeTe\ strongly suggests that the superconducting state (SC3) in the $x\gtrsim0.30$ region has a different mechanism from that in SC2.   
Indeed, the full-gap superconductivity, which is quite different from anisotropic superconductivity found in the SC2 region, has been reported by the scanning tunneling spectroscopy in optimally substituted \FeSeTe \ \cite{hanaguri2010unconventional}.
Although Raman spectroscopy revealed that in FeSe$_{0.4}$Te$_{0.6}$, the strength of electron-phonon coupling is insufficient to give \tc \ $= 14$\,K \cite{wu2020superconductivity}, no significant $(\pi,\pi)$ antiferromagnetic fluctuations are detected in the NMR experiments \cite{arvcon2010coexistence}, which is consistent with the recent high-pressure study in \FeSeTe \ showing that the pressure-induced antiferromagnetic order fades away above $x \sim 0.14$ \cite{Mukasa2020high}.
These results can preclude that the superconducting dome at $x\gtrsim0.30$ is associated with magnetic fluctuations and further support that the enhancement of critical temperature in the SC3 region comes from the nematic quantum-critical fluctuations observed in our elastoresistivity measurements.

Although several theories have shown that nematic fluctuations can enhance \tc, most of these theories consider purely electronic systems, which do not include the coupling to the underlying lattice inevitably present in real materials. 
Recently, however, it has been pointed out that this nemato-elastic coupling plays a crucial role in the nematic quantum criticality \cite{paul2017lattice}.
Through coupling to the lattice, the divergence of the correlation length at the nematic QCP is restricted only along the two high-symmetry regions, and the criticality can be cut off.
Therefore, the strength of nemato-elastic coupling is an important parameter at the nematic QCP, and this is closely related to the parameter $r_0 = (T_{\rm s}-T_{\rm 0})/T_{\rm F}$, where $T_{\rm F}$ is the Fermi temperature.
According to this theory, in FeSe-based materials with small Fermi energy, the effect of nemato-elastic coupling can be particularly significant compared to other iron-based superconductors. 
Indeed, our results show that the sign change of the Curie-Weiss temperature \tw\ estimated from the elastoresistivity data above \ts\ is shifted considerably from the thermodynamic QCP where \ts\ $\to 0$ in \FeSeTe. 
In contrast, however, no significant shift is found in our elastoresistivity data for \FeSeS. 
This apparent difference between \FeSeS\ and \FeSeTe\ may be related to the difference in the slopes of \ts\ with substitution near the QCPs; in \FeSeTe\ the sign change of \tw\ occurs at $x\approx 0.38$ where \ts\ $\approx 33$\,K, whereas \ts\ in \FeSeS\ changes steeply from 38 to 0\,K between which \tw\ changes its sign.
These results suggest that the nemato-elastic coupling is important to discuss the nematic quantum criticality in both systems. 

The recent theory predicts that the enhancement of \tc\ near the nematic QCP is expected only when the nemato-elastic coupling parameter $r_0$ is much smaller than the phenomenological ratio $(U/V)^2$, where $U$ and $V$ are the nematic and magnetic pairing interactions, respectively \cite{labat2017pairing}.
This can qualitatively account for the distinct difference between the absence and the presence of \tc \ dome around the nematic QCPs of \FeSeS \ and \FeSeTe. 
Although our observation of identically diverging nematic susceptibility around the two nematic QCPs (Fig.\,\ref{NS_QCP}) implies that the nematic interaction $U$ is similar in \FeSeS \ and \FeSeTe, the spin interaction term $V$ is considered quite different between them, as revealed by NMR measurements \cite{wiecki2018persistent,arvcon2010coexistence}.  
Namely, \FeSeS \ with no enhancement of \tc\  exhibits relatively strong spin fluctuations leading to a small $(U/V)^2$ parameter, which cannot satisfy the $r_0\ll (U/V)^2$ relation, whereas \FeSeTe, in which no significant antiferromagnetic fluctuations are found and thus a larger $(U/V)^2$ is expected, exhibits a clear superconducting dome near the thermodynamic nematic QCP. 

We point out that the two superconducting domes (SC2 and SC3) studied here may have some similarities with the phase diagram of hole-doped high-\tc \ cuprate superconductors under high magnetic fields, which also has two peaks in underdoped and slightly overdoped regions \cite{ramshaw2015quasiparticle}.
While the center of one dome locates near the endpoint of short-range antiferromagnetic order, the other with higher \tc \ is around the critical doping at which the enigmatic pseudogap phase terminates.
Recent studies show that significant electronic anisotropy develops inside the pseudogap phase \cite{daou2010broken,Lawler2010intra}, and there is evidence for enhanced nematic fluctuations at its critical point \cite{ishida2020divergent,auvray2019nematic}.
This similarity in the systems with quite different electronic structures may imply that nematic fluctuations can enhance superconductivity more strongly than previously thought, which stimulates further investigation.

Before concluding, we briefly mention that the present results are also intriguing from the topological aspects of \FeSeTe. 
Recent angle resolved photoemission spectroscopy revealed that FeSe$_{0.45}$Te$_{0.55}$ exhibits a topological superconducting state with the Dirac-type semimetallic bands above the Fermi level \cite{zhang2018observation,zhang2019multiple}.
Such topological bands in the bulk can couple to the nematic quantum fluctuations, and their interplay may host unique quantum phenomena which deserves future studies.  

\section{CONCLUSIONS}
In summary, the present systematic elastoresistivity measurements in \FeSeTe \ single crystals provide strong evidence for the nonmagnetic pure nematic quantum critical point accompanied by the superconducting dome. 
The enhancement of critical temperature in this material can be ascribed to the quantum critical fluctuations of the electronic nematic phase, which may offer a new route to high-temperature superconductivity. 

\section*{Acknowledgements}
We thank fruitful discussion with T. Hanaguri, H. Kontani, and I. Paul. This work was supported by Grant-in-Aid for Scientific Research (KAKENHI) on Innovative Areas ``Quantum Liquid Crystals" (No.\ JP19H05824), for Transformative Research Areas (A) “Condensed Conjugation” (No.\ JP20H05869), and by KAKENHI (Nos.\ JP20H02600, JP20K21139, JP19H00649, JP19J12149, JP19K22123, JP18KK0375, JP18H01853 and JP18H05227) from Japan Society for the Promotion of Science, and CREST (No.\ JPMJCR19T5) from Japan Science and Technology (JST).

\bibliography{ref.bib}


\section*{Methods}
\subsection*{Single crystals}
Single crystals of \FeSeS \ and \FeSeTe \ were grown by the chemical vapor transport technique \cite{matsuura2017maximizing,Mukasa2020high}.
The samples of \FeSeTe \ measured in this study are from the same batches used in Ref.~\cite{Mukasa2020high}.
Fe, Se, and S (Te) powders were mixed with AlCl$_{3}$ and KCl transport agents and sealed in evacuated quartz ample.
Temperatures of the source and sink sides were set at 420\(^\circ\)C and 250\(^\circ\)C for \FeSeS \ and \FeSeTe \ with $x \lesssim 0.25$, and 620\(^\circ\)C and 450\(^\circ\)C for \FeSeTe \ with $x \gtrsim 0.25$, respectively. 

The actual Te and S compositions were determined for each sample before the elastoresistivity measurements, 
by the $c$-axis length measured by X-ray diffraction  (see Fig.\,\ref{PD}\FigCap{a}). 
For all the samples, homogeneous distributions of chalcogen ions were confirmed by energy-dispersive X-ray spectroscopy (see Fig.\,\ref{figS1}). \\

\subsection*{Elastoresistivity measurements}
For the systematic measurements of nematic susceptibility, we adopt the elastoresistivity measurement technique using the piezoelectric device.
In this technique, we measure the strain-induced in-plane resistivity anisotropy.
The experimental setup is shown in Fig.\,\ref{ER}\FigCap{a}.
The samples are cut into square shapes, and the resistivity along the $x$ and $y$ directions ($\rho_{xx}$ and $\rho_{yy}$) are measured by the Montgomery method.
One advantage of this method is that we can measure both $\rho_{xx}$ and $\rho_{yy}$ in a given sample.
To discuss the systematic dependence of the magnitude of the nematic susceptibility, we set the lateral sample size  approximately fixed to $250\,\mu$m $\times 250\,\mu$m to minimize the possible size dependence.

After making the electrical contacts on the prepared samples, we glued them on the piezo stacks.
The strain was transmitted to the sample via orthorhombic distortion of the device and controlled in-situ by applying the voltage to the piezo stack.   
The amount of the strain $\varepsilon_{xx}$ was measured by a strain gauge attached on the backside of the device, and the orthogonal strain $\varepsilon_{yy}$ was calculated by the Poisson's ratio of piezo stacks calibrated beforehand.


\section*{Supplementary Information}
\renewcommand{\thefigure}{S\arabic{figure}}
\setcounter{figure}{0}

\subsection{Characterization of single crystals}
\begin{figure}[t]
	\centering
	\includegraphics[width=\linewidth]{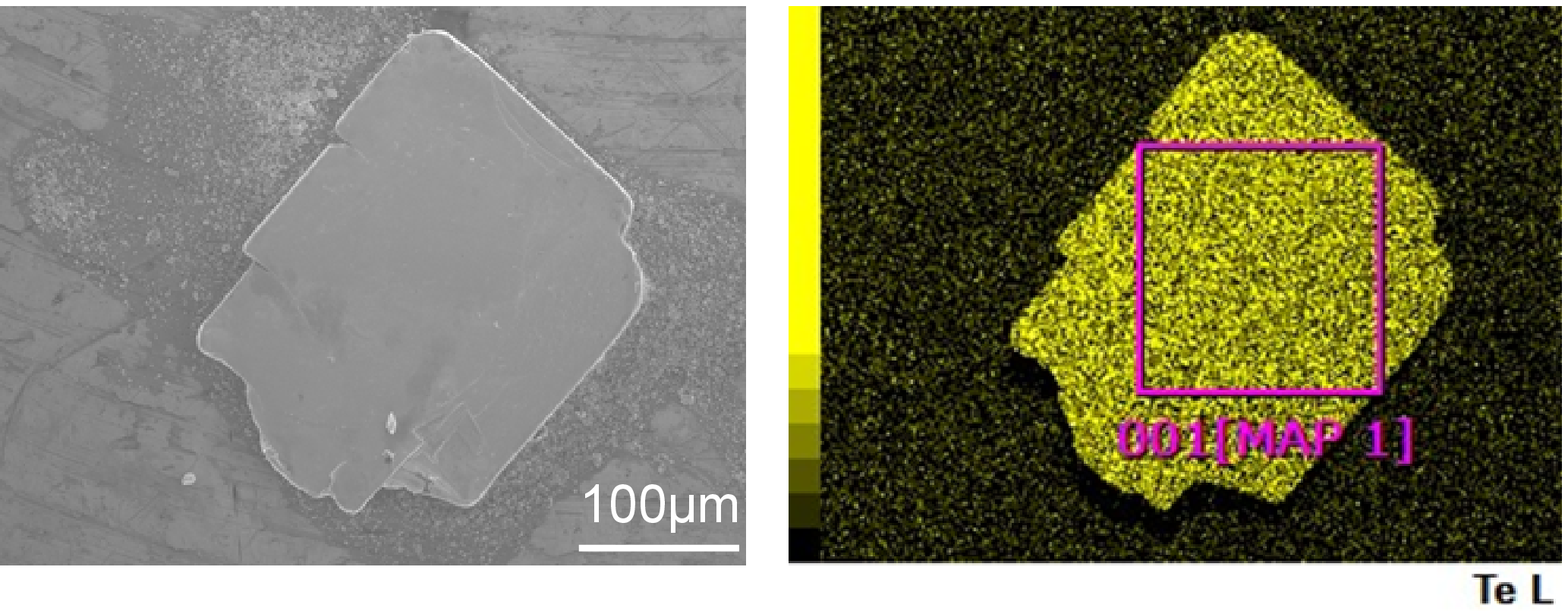}
	\caption{Energy dispersive X-ray spectroscopy for a \FeSeTe \ single crystal.
	Left panel shows the scanning electron microscope image of a $x=0.34$ sample, and right panel represents the Te distribution inside the sample as a color plot.
	}
	\label{figS1}
\end{figure}

Before measuring the elastoresistivity, we characterized the actual S and Te compositions in \FeSeS\ and \FeSeTe\ by using energy-dispersive X-ray spectroscopy (EDX) and X-ray diffraction.
The representative results of EDX analysis for \FeSeTe \ single crystals are shown in Fig.\,\ref{figS1}.
Homogeneous distributions of Te ions were seen in all the samples, and the average Te concentration over the square marked in Fig.\,\ref{figS1} (MAP1) is consistent with that determined by X-ray diffraction. 

\subsection{Comparisons between $B_{1g}$ and $B_{2g}$ nematic susceptibilities in \FeSeTe}

\begin{figure}[t]
	\centering
	\includegraphics[width=\linewidth]{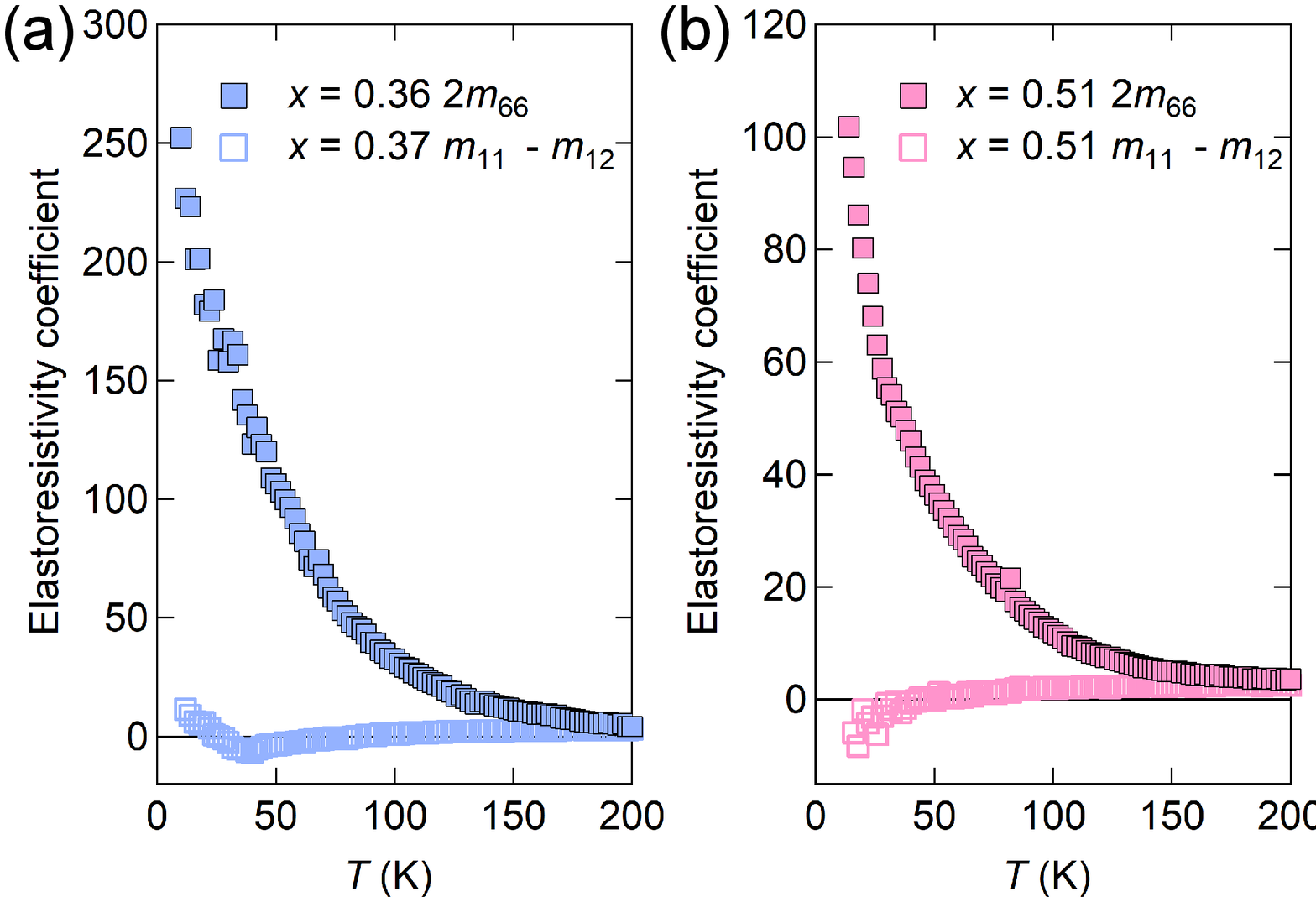}
	\caption{Two nematic susceptibilities of \FeSeTe \ near the nematic QCP.
	\FigCap{a,b} Temperature dependence of $m_{11}-m_{12}$ and $2m_{66}$ for \FeSeTe \ with $x\sim0.36$ (\FigCap{a}) and $x=0.51$ (\FigCap{b}).
	}
	\label{figS2}
\end{figure}

Figure\,\ref{figS2} represents the two elastoresistivity coefficients $m_{11}-m_{12}$ and $2m_{66}$ for $x\sim0.36$ with \tw \ $\sim$ 0\,K and $x=0.51$ near the endpoint of \ts.
In both samples, $m_{11}-m_{12}$ has subtle temperature dependence, which cannot be explained by the contamination of $2m_{66}$ due to the misalignment of the strain direction.
The magnitude of $m_{11}-m_{12}$ is much smaller than that of $2m_{66}$, evidencing that the nematic QCP in \FeSeTe \ has a $B_{2g}$ Ising character.

\subsection{Elastoresistivity coefficient $2m_{66}$ in \FeSeS}
\begin{figure}[ht]
	\centering
	\includegraphics[width=\linewidth]{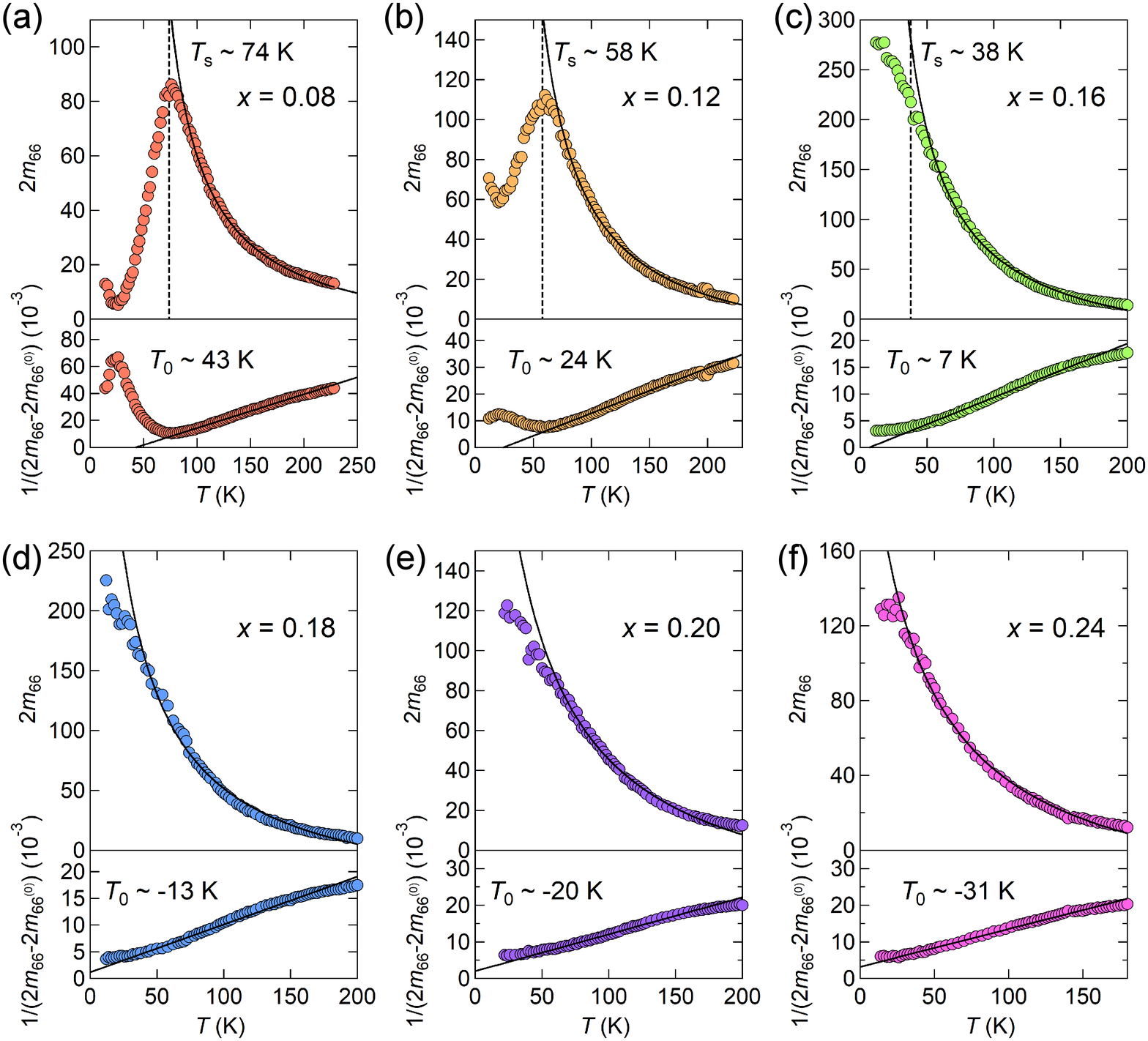}
	\caption{$B_{2g}$ nematic susceptibility in \FeSeS.
	\FigCap{a-f}  Temperature dependence of $2m_{66}$ in \FeSeS \ single crystals with $x=0.08$ (\FigCap{a}), $0.12$ (\FigCap{b}), $0.16$ (\FigCap{c}), $0.18$ (\FigCap{d}), $0.20$ (\FigCap{e}), and $0.24$ (\FigCap{f}).
	Black lines represent the results of Curie-Weiss fitting. 
	\ts \ of each sample is shown in vertical dashed line. Each bottom panel shows the Curie-Weiss analysis and the obtained value of \tw.
	The data shown here together with the data in Fig.\,\ref{NS} are used to construct the color plot of Fig.\,\ref{PD}\FigCap{b}.}    
	\label{figS3}
\end{figure}

We have also performed the elastoresistivity measurements in \FeSeS \ single crystals with 6 different sulfur compositions covering inside and outside of the nematic phase.
The results of the temperature dependence of $2m_{66}$ are shown in Fig.\,\ref{figS3}.
For $x\leq0.16$, $2m_{66}(T)$ displays the Curie-Weiss behavior toward the structural transition temperature.
This Curie-Weiss temperature dependence also holds after the nematic phase disappears at $x\sim0.17$, but near the nematic quantum critical point, the downward deviation is apparent at low temperatures, which possibly comes from the pronounced sensitivity of disorder to the quantum criticality \cite{kuo2016ubiquitous}.
From the Curie-Weiss analysis, \tw \ is found to change its sign across the nematic phase boundary (between $x=0.16$ and $x=0.18$), indicating the nematic quantum critical point near $x=0.17$.

\end{document}